\documentclass[aps,pre,
twocolumn,
superscriptaddress,floatfix]{revtex4-2}

\usepackage[margin=0.9in]{geometry}
\usepackage[english]{babel}
\usepackage[utf8]{inputenc}
\usepackage{subfiles}
\usepackage{graphicx}
\usepackage{caption}
\usepackage{subcaption}
\usepackage{soul}
\usepackage{float}
\usepackage{amsmath}
\usepackage{bbm}
\usepackage{xcolor}
\usepackage[T1]{fontenc}

\begin{document}

\title{Opinion dynamics under large-scale exogenous events}

\author{Gerard Argany Herrera}
\affiliation{Departament de Física de la Matèria Condensada,
Universitat de Barcelona, Mart\'i i Franqu\`es 1, 08028 Barcelona, Spain}

\author{Lucila G. Alvarez-Zuzek}
\affiliation{Complex Human Behaviour Laboratory,
Fondazione Bruno Kessler, Trento, Italy}

\author{Oriol Artime}
\email{oartime@ub.edu}
\affiliation{Departament de Física de la Matèria Condensada,
Universitat de Barcelona, Mart\'i i Franqu\`es 1, 08028 Barcelona, Spain}
\affiliation{
Universitat de Barcelona Institute of Complex Systems (UBICS),
Universitat de Barcelona, Mart\'i i Franqu\`es 1, 08028 Barcelona, Spain}

    \begin{abstract}
         Large-scale events such as political scandals or misinformation campaigns can abruptly shift the collective opinion of a population and reshape the dynamics in ways that standard local-interaction models of sociophysics do not capture. Here, we investigate the dynamics of opinion formation when a group of interacting agents is subjected to sudden, macroscopic exogenous shocks. Our framework is general and applies to arbitrary opinion dynamics models; to illustrate its scope, we consider the voter model and variations thereof. We verify that the mathematical description is excellent in describing the first-passage properties to consensus of the voter model on the complete graph and on complex networks. Remarkably, we identify the conditions under which shocks can be advantageous or detrimental to reach consensus. For the case of the all-to-all noisy voter model under exogenous events, we unveil a rich phase diagram characterized by novel trimodal and asymmetric bimodal states. Ultimately, this work provides a rigorous yet flexible stochastic-process framework for understanding the out-of-equilibrium behavior of opinion dynamics under recurrent exogenous stressors.
    \end{abstract}
    \maketitle

{\textbf{Significance statement: Collective behavior is often understood as emerging from the accumulation of local interactions among individuals. However, real-world social, biological, and technological systems are also shaped by sudden, collective external perturbations, such as political scandals, misinformation campaigns, or other large-scale events, which can abruptly alter the trajectory of a system. Here, we develop a general stochastic framework for studying the interplay between endogenous opinion dynamics and exogenous shocks. Applying it to voter-model dynamics, we show how such shocks can either accelerate or hinder consensus formation and can generate qualitatively new collective states, including trimodal and asymmetric bimodal regimes. More broadly, our results demonstrate that rare macroscopic perturbations can fundamentally reshape collective dynamics and highlight the need for mathematical frameworks that incorporate both local interactions and external forcing. Such tools may be essential for understanding, predicting, and ultimately steering the emergence of collective states in complex systems}}

    \section{Introduction}
    \label{sec:introduction}

    In the last decades, the theory of complex systems has provided a convenient framework for understanding a wide variety of real-world collective phenomena~\cite{artime2022origin}. The shared characteristic to these phenomena is that they occur in many-body systems, where properties at a large scale emerge due to the interactions of the elements at lower scales~\cite{anderson1972more}. This turned out to be ubiquitous across disciplines~\cite{castellano2009statistical, stanley2002self, de2026decoding, strogatz2022fifty}, and it has been addressed, with a toolkit developed in statistical physics, information theory and network science, phenomena as diverse as human mobility and migration~\cite{barbosa2018human}, cultural dissemination~\cite{axelrod1997dissemination}, language evolution~\cite{abrams2003modelling}, collective decision making~\cite{vicsek2012collective} and opinion formation~\cite{caldarelli2026physics}, among others.

    The study of macroscopic dynamics of social systems have flourished under this umbrella, crystallizing in the so-called sociophysics~\cite{sen2014sociophysics, schweitzer2018sociophysics, galam2012sociophysics}. Its research agenda aims to relate individual decisions, choices and behaviors to large-scale social phenomena. Among these, opinion models have been proposed to better understand consensus formation, polarization, echo chambers, etc~\cite{caldarelli2026physics}. In them, individuals are represented by agents whose opinion evolves according to simple interaction rules that capture the mechanism of social influence under study. 
    
    Despite the rich phenomenology captured by these models, a significant gap remains when trying to connect agent-level interactions to sudden, system-wide changes in public opinion. The exposure to large-scale exogenous shocks, such as political scandals, coordinated misinformation campaigns, terrorist attacks, economic crises, or major scientific announcements can abruptly reshape the opinion landscape~\cite{baker2001patriotism, lipset1967party, collingwood2020change, martin2022metanoia}. These events may polarize society, cause a global shift to the minority opinion, or reinforce pre-existing opinions of a significant fraction of the population, in timescales that nontrivially intertwine with those of local-based social mechanisms. However, in classical opinion dynamics models~\cite{sznajd2021review, weisbuch2002meet, rainer2002opinion, galam2002minority}, the mechanisms of social influence are implemented such that there is at most one opinion change per interaction. Thus, we currently lack models that incorporate abrupt macroscopic opinion changes coupled to local mechanisms of social influence, evincing a gap between traditional assumptions behind opinion dynamics models in sociophysics and the large-scale perturbations observed in empirical scenarios.

    To fill this gap, we formulate a flexible framework to extract qualitative and quantitative predictions of scenarios in which an arbitrary local-based dynamics is coupled to large-scale exogenous shocks. The shocks have tunable properties, in order to account for different empirically-motivated cases. We illustrate the excellent performance of the framework using a canonical model of sociophysics, the voter model, as the reference local-based model. The voter model is a bare-bones, analytically tractable opinion model that sheds light into the role played by pair-wise imitation in shaping consensus formation. There exist plenty of generalizations that account for other mechanisms of social influence, such as idiosyncrasy, zealotry and conformity, among others; see~\cite{redner2019reality, jkedrzejewski2019statistical} for reviews. To showcase the flexibility of the framework, we apply it to one of these generalizations of the voter model, i.e., its noisy version, also known as the Kirman model~\cite{kirman1993ants}. Agents interacting in a complete (all-to-all) graph is the most mathematically transparent case, yet this simplistic case is overcome thanks to a pair approximation that helps us unravel the influence of the network of interactions among agents on the coupled model. Typical equilibrium and non-equilibrium properties are explored in all cases: stationary distribution, full-time solution and first-passage statistics to consensus states. 

    The remainder of the article is as follows. We first introduce the mathematical framework in the voter model, detailing how to compute the different quantities of interest. A time-discrete version of the dynamics is used, and we leave in the appendix a treatment of the continuous-time scenario. We then extend the treatment of exogenous events for the voter model defined in complex networks. Finally, the all-to-all case for the noisy voter model is presented, discussing the rich phase diagram that emerges due to the external shocks. To close, we draw the conclusions.

    \section*{Exogenous shocks in the voter model}
    \label{sec:vm}

    As anticipated, exogenous shocks are events that are coupled to some baseline local dynamics. We here introduce first the discrete-time voter model, acting as baseline dynamics, to then explain how to incorporate the shocks.

    We take a population of $N$ agents, where each agent is represented by a node $i$ holding a discrete and binary opinion denoted by the state variable $s_i \in \{-1,1\}$, also known as its opinion. These agents are connected through an undirected network defining their possible interactions~\cite{newman2018networks}. The voter dynamics goes as follows. At every time step, a random agent and one of its neighbors are selected, and the former copies the opinion of the latter. This is repeated \textit{ad infinitum}. To track the macroscopic state of the entire system as these microscopic imitations occur, we define the global magnetization
\begin{equation}
  m = \frac{1}{N}\sum_{i = 1}^N s_i \in [-1,1].
\end{equation}
    The magnetization serves as the primary order parameter for the system. A state where $m=0$ indicates a balanced fragmentation (an equal number of agents holding $+1$ and $-1$ opinions), whereas $m=\pm 1$ indicates total consensus of one of the two opinions. With every single opinion change, the magnetization jumps by discrete steps of $\Delta m = \pm 2/N$ and the time increases by $\Delta t = 1/N$.

    The repeated application of the update rules generates a stochastic random walk in the magnetization space where the jumping rates/probabilities are state dependent. The so-called absorbing states are reached when every agent shares the same opinion ($m = +1$ or $m = -1$) and no further updates can modify the system. This naturally motivates the study of the first-passage time (FPT) to consensus~\cite{artime2018first}. In an unconditional scenario, the mean first-passage time (MFPT) is defined as the expected time it takes for the system to hit either of the absorbing states from an initial starting magnetization $m_0$. In the conditioned scenario, only the trajectories that hit one of the two consensus states before the other are considered.

    The discrete-time voter model admits a Markov chain representation whose states are defined by the number $n \in \lbrace 0, 1, \ldots, N \rbrace$ of agents holding opinion $1$~\cite{pickering2015solution}. The corresponding magnetization is given by $m = 2 n / N-1$, and we use $n$ and $m$ interchangeably. At each time step, $n$ changes by $\pm1$, i.e., 
\begin{equation}
    n(t+1) = n(t) + \Delta n(t), 
\end{equation}
    where $\Delta n$ is a random variable. The transition probabilities are computed by multiplying the probabilities of choosing an agent of opinion $1$ and then another of opinion $-1$, or vice versa,
\begin{align}
    & P(\Delta n(t) = 1 \,|\,n(t) = j) = \frac{j}{N}\frac{N-j}{N-1} \equiv p_j, \\
    & P(\Delta n(t) = -1 \,|\,n(t) = j) = \frac{N-j}{N-1} \frac{j}{N} \equiv p_j.
\end{align}
    The probability of not observing an opinion change is
\begin{equation}
    P(\Delta n(t) = 0 \,|\,n(t) = j) = 1-2p_j.
\end{equation}
    Then, the master equation for $a^{(t)}_j \equiv P(n(t) = j)$ is
\begin{equation}
    a^{(t+1)}_j = a^{(t)}_{j+1} p_{j+1} + a^{(t)}_j\left( 1-2p_j \right) + a^{(t)}_{j-1} p_{j-1}.
\end{equation}
    Equivalently, as a matrix equation,
\begin{equation}
    \begin{bmatrix} a_0^{(t+1)} \\ a_1^{(t+1)} \\ \vdots \\ a_N^{(t+1)} \end{bmatrix}   
    = \begin{bmatrix}
      1 & p_1 & 0 & \cdots & 0 \\
      0 & 1 - 2p_1 & p_2 & \cdots & 0 \\
      0 & p_1 & 1 - 2p_2 & \cdots & 0 \\
      \vdots & \vdots & \vdots & \ddots & \vdots \\
      0 & 0 & 0 & \cdots & 1
    \end{bmatrix} 
    \begin{bmatrix} a_0^{(t)} \\ a_1^{(t)} \\ \vdots \\ a_N^{(t)} \end{bmatrix},
\end{equation}
or
\begin{equation}
    \label{eq:baseline_matrix1}
    \mathbf{a}^{(t+1)} = \mathbf{P} \cdot \mathbf{a}^{(t)},
\end{equation}
whose formal solution is
\begin{equation}
    \label{eq:baseline_matrix2}
    \mathbf{a}^{(t)} = \mathbf{P}^t \cdot \mathbf{a}^{(0)}.
\end{equation}
    Thus, given an initial condition $\mathbf{a}^{(0)}$, the probability mass function at any time is accessible just by a simple matrix exponentiation. 

\begin{figure}[ht]
    \centering
    \includegraphics[width=1\linewidth]{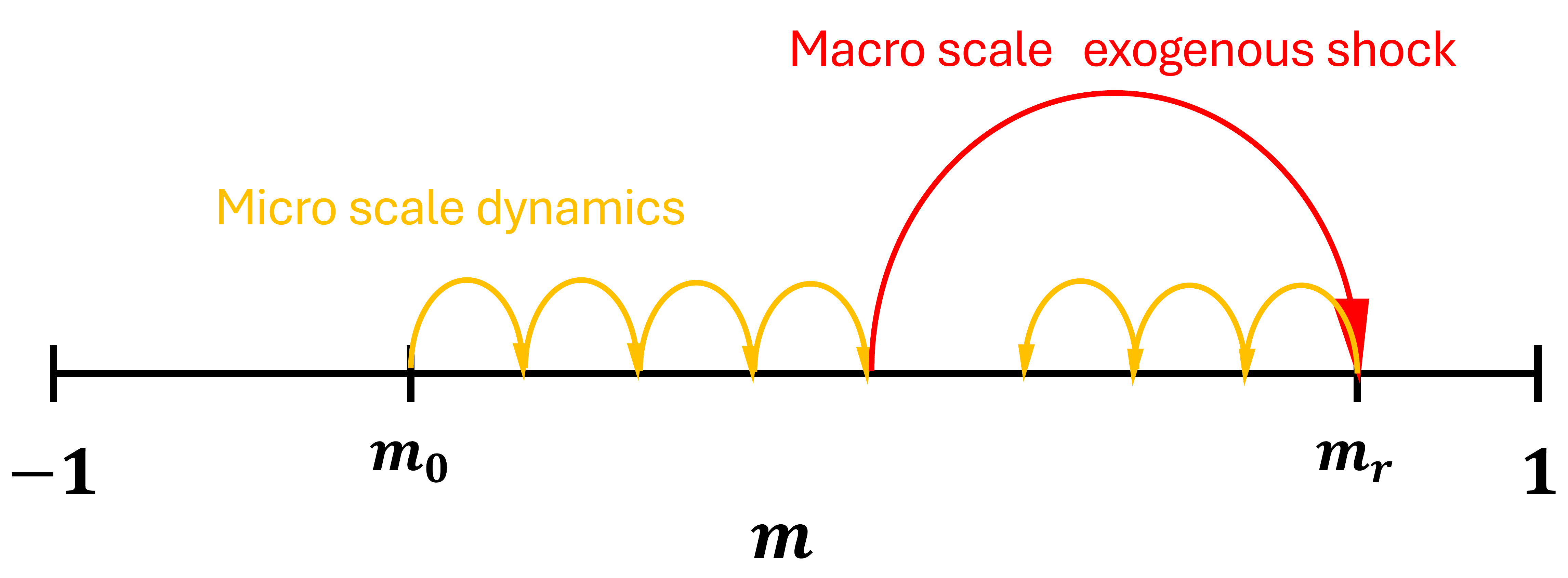}
    \caption{\textbf{Sketch of the macroscopic shifting dynamics}. Random walk steps across the magnetization domain $m \in [-1, 1]$, randomly interrupted by and exogenous events that force the system to jump to a state $m_r$ drawn from the distribution $\psi(m_r)$.}
    \label{fig:example}
\end{figure}

    We introduce exogenous shocks as events occurring with probability $r$ that trigger a collective shift of the system by instantaneously driving it to a macroscopic state of magnetization $m_r$, drawn from a distribution function $\psi(m_r)$. This means that during these events, in general $|\Delta m| > 2/N$, or $|\Delta(n)| > 1$, while the time increment remains $\Delta t = 1/N$. With the complementary probability $1-r$ the system evolves following the baseline dynamic. A sketch of this composite dynamics in the magnetization space is presented in Fig.~\ref{fig:example}.

    Mathematically, the transition matrix of the composite process $\mathbf{P}_r$ is inevitably different to the baseline process' $\mathbf{P}$. However, one can still write 
\begin{equation}
    \mathbf{P}_r = (1-r)\mathbf{P} + r\mathbf{R},
    \label{eq:pr}
\end{equation}
    and apply it in Eqs.~\eqref{eq:baseline_matrix1} and \eqref{eq:baseline_matrix2} to have the full-time evolution of the opinion dynamics including the exogenous events. The properties of the shocks are encoded in the matrix $\mathbf{R}$. For instance, for the simple case in which the shocks drive the magnetization always to a reference value $m_r$, corresponding to $n_j$, $\mathbf{R}$ is $0$ everywhere except for $1$s in the $j$th row. For the complete opposite scenario in which an exogenous shock drives the system to any possible state with equal probability, then $\mathbf{R}$ is a constant matrix whose entries are all $1/N$. 

    In the classical voter model, consensus states are absorbing. Exogenous events, though, permit the system escape those with probability $r$, so the absorbing nature is lost and ergodicity is restored. This key modification creates a non-equilibrium steady states (NESS) with non-trivial maxima and minima that are worth exploring as a function of $r$, $\mathbf{R}$ and potential parameters of the baseline model.

    First-passage times to consensus for the voter model are well known. Thus, for the sake of comparison, it is instructive to shed light on the role played by the exogenous shocks on metrics related to FPT. To be precise, we address separately the cases of the unconditional MFPT, where we do not distinguish between reaching $m=+1$ or $m=-1$, and of the conditioned MFPT where we only consider trajectories that reach the $m = 1$ consensus state. 
    
    To find the mean first-passage time in the time-discrete setting, we exploit the theory of absorbing Markov chains. Let us consider the matrix $\mathbf{Q_r}$, taken as the $\mathbf{P}_r$ matrix, Eq.~\eqref{eq:pr}, without the consensus states. Then, we construct the fundamental matrix,
\begin{equation}
    \mathbf{N_r} = \sum_k^\infty \mathbf{Q_r}^{k} = (\mathbf{I} - \mathbf{Q_r})^{-1}.
\end{equation}
    It is known that the MFPT to consensus when starting from some state $m_0$ is~\cite{kemeny1969finite}
\begin{equation}
    \mathbf{\tau}(r) = \mathbf{N_r} \cdot \mathbf{\mathbbm{1}},
    \label{eq:MFPT}
\end{equation}
    where $\mathbbm{1}$ is a column vector of $1$s and $\mathbf{\tau}(r)$ is a column vector in which at each entry we have the MFPT starting from the corresponding $m_0$. 

\begin{figure}[t!]
    \centering
    \includegraphics[width=1\linewidth]{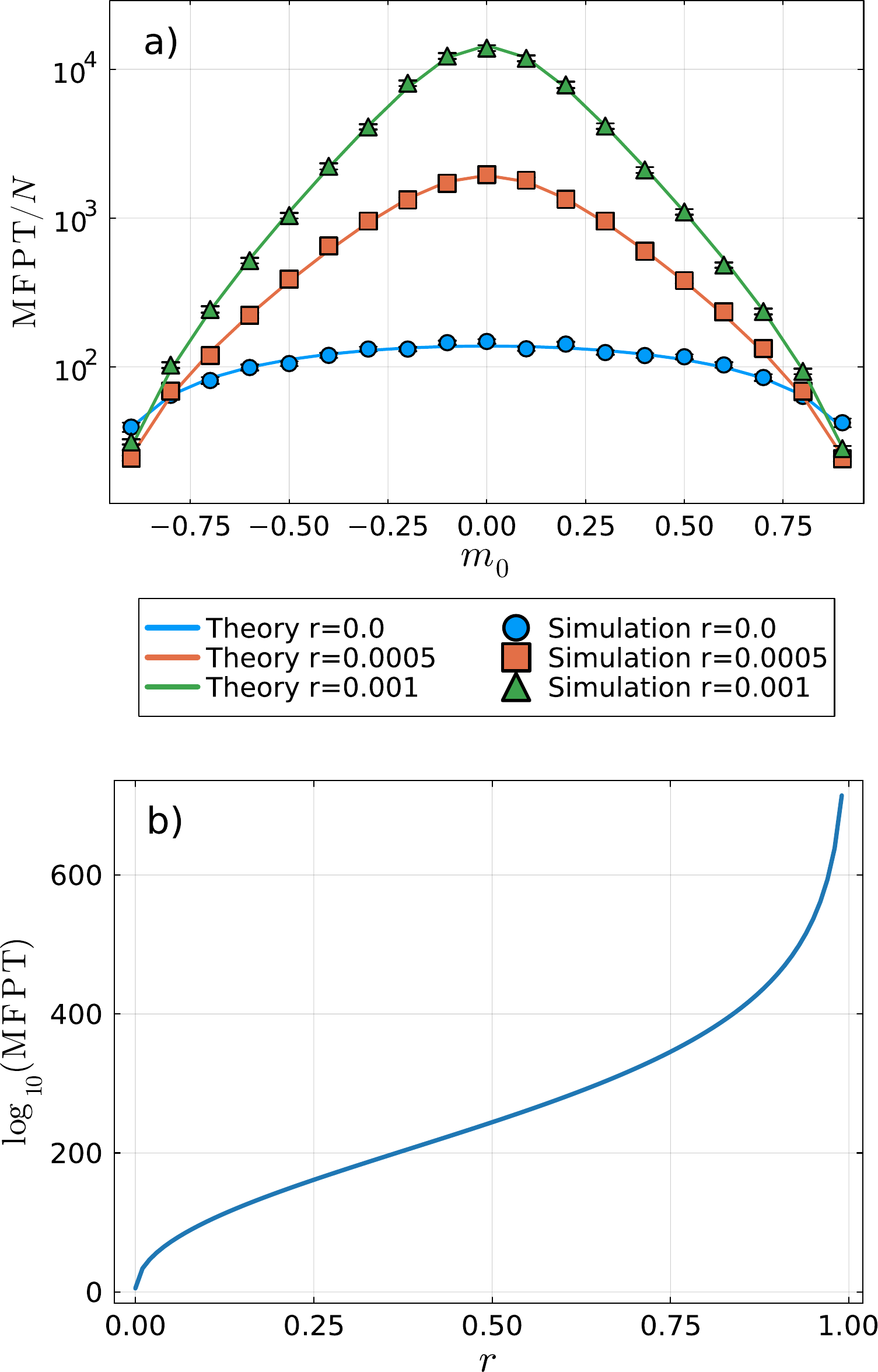}
    \caption{\textbf{Unconditional mean first-passage time to any consensus state}. In a), we compare the exact discrete theory from Eq.~\ref{eq:MFPT} (lines) with Monte Carlo simulations (symbols) as a function of the initial magnetization $m_0$, considering $N = 200$. Different curves correspond to different values of the probability of exogenous events $r$. In b), we display the MFPT to consensus as a function of $r$, for fixed $m_0 = 0$ and $N = 500$. The rapid growth illustrates how frequent exogenous shocks can severely impede the formation of consensus. In all cases, we use $\delta$-type shocks to the initial magnetization $m_0$.}
    \label{fig:fig2}
\end{figure}

    For illustrative purposes, let us take the voter model departing from a magnetization $m_0$ and a shock distribution $\psi(m_r) = \delta_{m_r, m_0}$, i.e., the shocks drive the system always to a reference state $m_r$ that happens to be the initial one. The accuracy of the theoretical approach presented above is excellent, as it is seen in Fig.~\ref{fig:fig2}a). We note that, when the system is driven close to the state of equal coexistence of opinions, $m_0=0$, i.e., the farthest from consensus, the MFPT increases significantly higher as the rate of exogenous events $r$ does so. Actually, a very little variation in $r$ can have a massive influence on the MFPT. To quantify this, in Fig.~\ref{fig:fig2}b), we show how the maximum at $m_0 = 0$ scales with $r$, reaching up to $10^{700}$ for $r \approx 0.99$. Thus, computationally observing such arrivals to consensus is highly unlikely, revealing the advantage of the theoretical approach to estimate the statistics of these events.

    A closer inspection to the behavior of the mean first-passage time reveals that the strong delay in achieving consensus observed for $m_0 \approx 0$ when $r$ grows does not hold for all values of the initial magnetization. As can be seen in Fig.~\ref{fig:fig2}a), the MFPT curves for different $r$ cross at large values of $|m_0|$, meaning that, above a certain threshold $|m_0| > |m_c|$, increasing the rate of exogenous events leads to a decrease in the mean first-passage time. This behavior can be explained by the fact that, close enough to consensus, the number of trajectories that stray towards $m = 0$ is high enough so that returning to the initial $m$ brings the system closer to consensus on average. 
    
\begin{figure}[t!]
    \centering
    \includegraphics[width=1\linewidth]{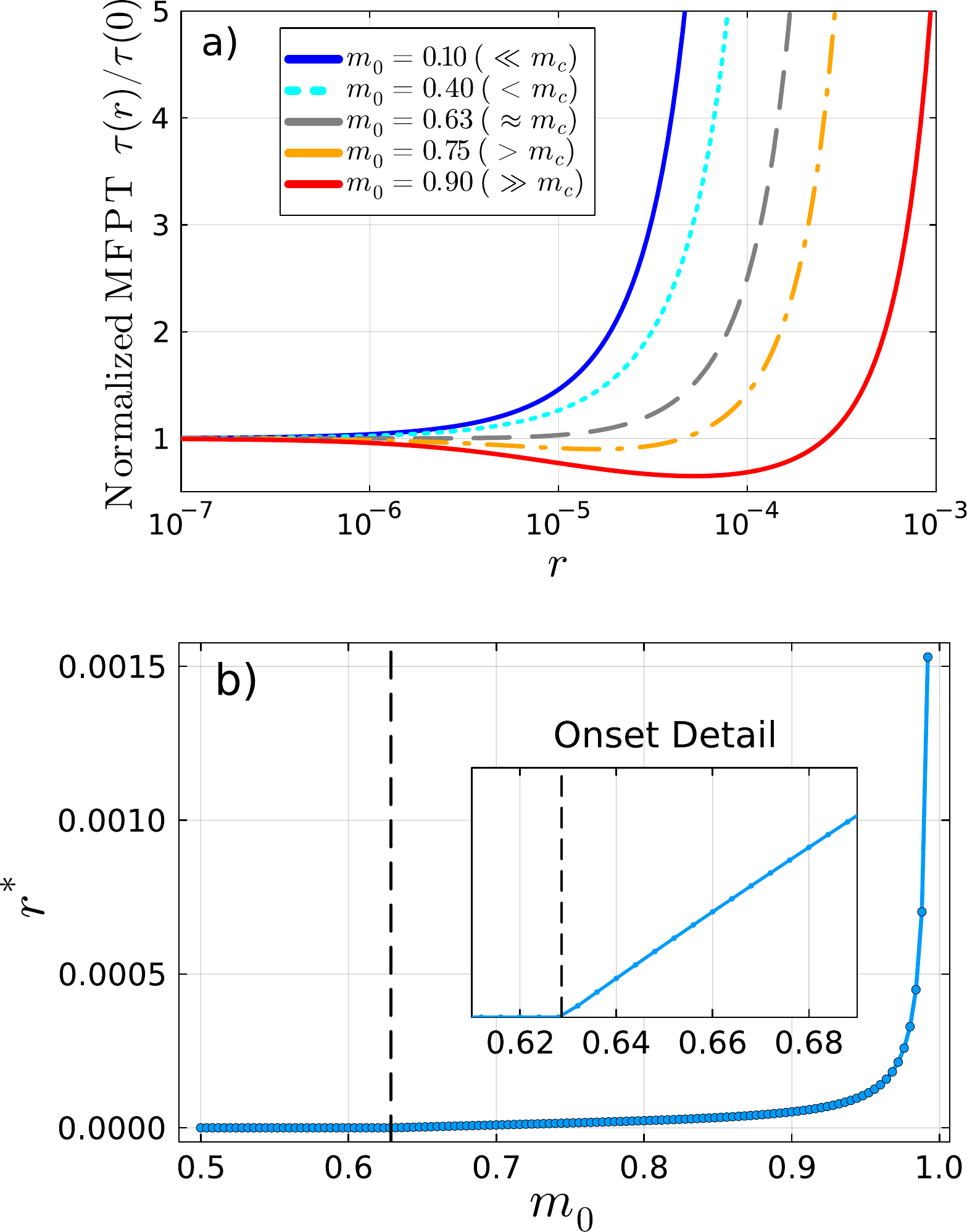}
    \caption{\textbf{Exogenous shocks can facilitate the arrival to consensus states}. In a), we show the normalized mean first-passage time to consensus as a function of shock probability $r$. Below the critical threshold ($m_0 < m_c$), the exogenous shocks delay the reach of consensus. Conversely, above the threshold ($m_0 > m_c$), the curves presents a non-trivial minimum, revealing an optimal probability $r^*$ that minimizes the time to consensus. In b), we display the optimal rate $r^*$ as a function of $m_0$. In the inset, we zoom in around the transition at $m_c \approx 0.630$ (vertical dashed line, obtained from the theoretical value), marking the threshold between the region in which shocks inhibit consensus and the one in which shocks favor consensus. In all cases we consider $N = 500$ and $\delta$-type shocks to the initial magnetization $m_0$.}
	\label{fig:fig3}
\end{figure}

    We report the existence of a region of values of the initial magnetization $|m_0|$ for which the mean first-passage time becomes smaller than the reference value of the baseline $r=0$, i.e., due to the exogenous events one can accelerate the arrival to consensus; see Fig.~\ref{fig:fig3}a). Since $\tau(r) \gg \tau(0)$ when $r \to 1$ and $\tau(r) \to \tau(0)$ when $r \to 0$, if that region exist, it should present an absolute minimum in $\tau(r^*) / \tau(0)$ at some $r^*$. To find the critical values $|m_c|$ at which the optimal rate $r^*$ emerges, we exploit the condition~\cite{reuveni2016optimal}
\begin{equation}
    \text{Var} \left( T_{0}(m_c) \right) = \langle T_{0}(m_c) \rangle^2,
\end{equation}
    where $T_0(m_0)$ is the FPT at a rate of zero starting from $m_0$. We discover that exogenous shocks are expected to start favoring consensus around $m_0 = 0.630 \pm 0.002$. In Fig.~\ref{fig:fig3}b), we show the optimal rate $r^*$ as a function of $|m_0|$, finding an excellent agreement between theory and simulations. In the inset of Fig.~\ref{fig:fig3}b), we see that close to the point at which the minimization emerges, $r^*$ accepts a scaling relation
\begin{equation}
    r^* \sim (m-m_c)^{\beta} \qquad  \beta = 0.996 \pm 0.004.
\end{equation} 
    This fit is compatible with a linear relation. It remains as an open problem to derive this scaling analytically.

    To enrich the characterization of first-passage events under the exposure of exogenous shocks, we next focus in conditioned MFPTs. Those represent the times for trajectories that reach one of the two states of consensus without having before reached the other. Without loss of generality, we take the target consensus state to be $m=1$.

    To compute this MFPT theoretically, we apply a Doob h-transform~\cite{levin2026markov} to our composite matrix $\mathbf{P_r}$ (Eq.~\eqref{eq:pr}). Element-wise, we obtain
\begin{equation}
    \label{eq:hat_pij}
    \hat{P}_{ij} = \frac{h_j}{h_i} P_{ij},
\end{equation}
    where $h_i$ is the probability of ending at $m = 1$ having started from a magnetization $m_0$ corresponding to state $n_0 = i$. To find this $\mathbf{h}$, we make use of a theorem~\cite{kemeny1969finite} that states that
\begin{equation}
    \mathbf{h} = \mathbf{N_r} \cdot \mathbf{u},
\end{equation}
    where $\mathbf{u}$ is the vector that encodes the probabilities of jumping to $m = 1$ at the next time step, which in this case is
\begin{equation}
   \mathbf{u} = \begin{bmatrix} 0 \\ \vdots \\ 0 \\  p_{N-1}\end{bmatrix}.
\end{equation}
    Matricially, Eq.~\eqref{eq:hat_pij} can be written as
\begin{equation}
    \mathbf{\hat P_r} = \mathbf{D^{-1}}\mathbf{P_r}\mathbf{D}, \qquad \mathbf{D} \equiv \text{diag}(\mathbf{h}).
\end{equation}
    It is easy to prove that
\begin{equation}
    \mathbf{\hat N_r} = \mathbf{D^{-1}}\mathbf{N_r}\mathbf{D},
\end{equation}
    so our final conditioned MFPT is
\begin{align}
    \mathbf{\hat \tau_r} & = \mathbf{\hat N_r} \cdot \mathbf{\mathbbm{1}} \nonumber \\ 
    & = \mathbf{D^{-1}}\mathbf{N_r}\mathbf{D} \cdot \mathbf{\mathbbm{1}} \\
    & = \mathbf{D^{-1}}\mathbf{N_r} \mathbf{h}. \nonumber
\end{align}

\begin{figure}[t!]
    \centering
    \includegraphics[width=1\linewidth]{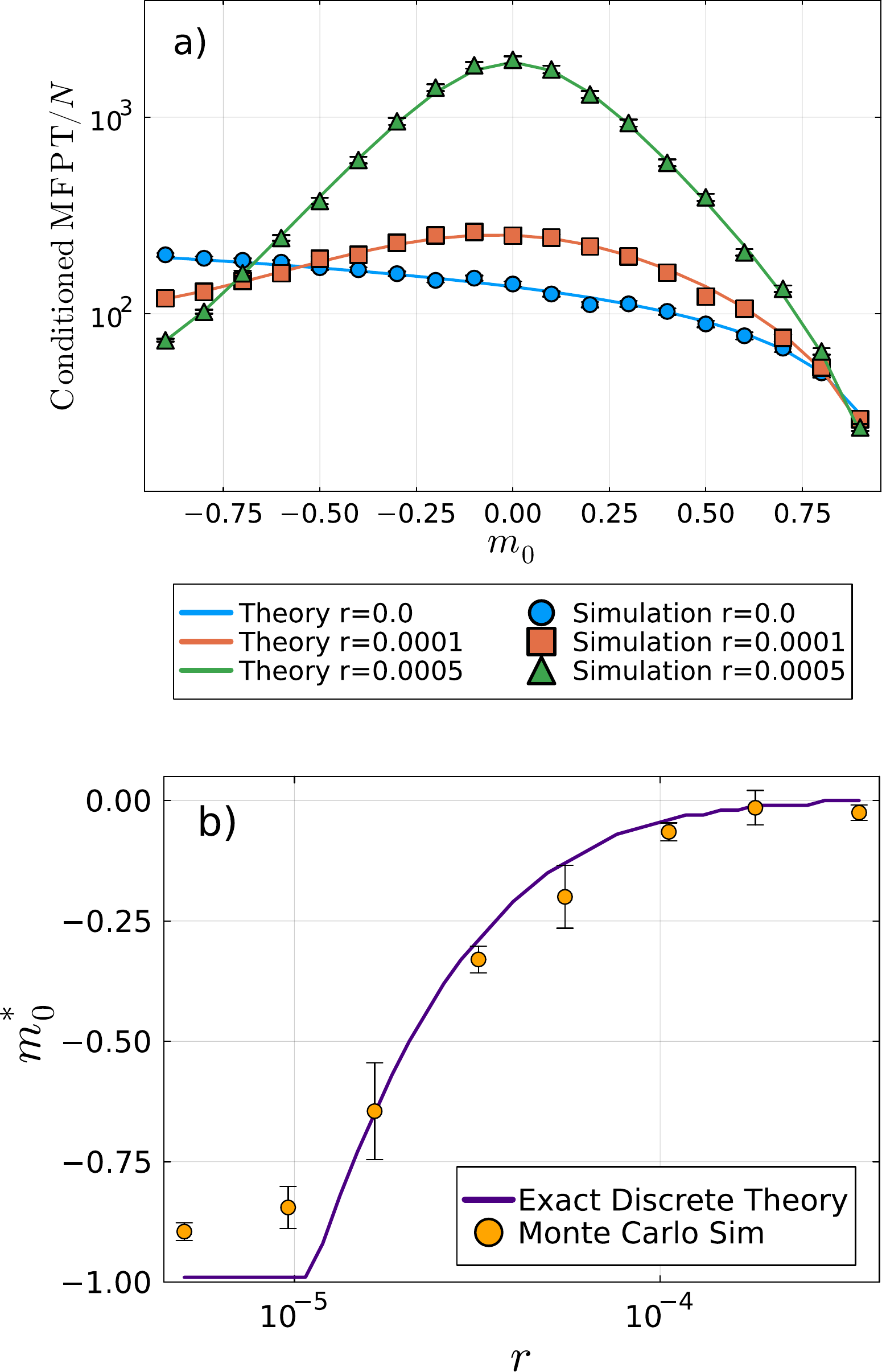}
    \caption{\textbf{Analysis of the conditioned mean first-passage time}. In a), we show the conditioned MFPT to reach consensus $m = +1$ without prior absorption at $m = -1$, as a function of $m_0$ across various shock probabilities $r$. In b), we present the position of the maximum conditioned MFPT ($m_0^*$) as a function of $r$. We observe that this maximum quickly shifts toward opinion coexistence ($m_0^* = 0$) as $r$ grows. In all cases, lines correspond to the exact discrete theory while markers correspond to simulated trajectories, and we take a system size $N = 200$. The shock events used here are of $\delta$-type to the initial magnetization $m_0$.}
	\label{fig:fig4}
\end{figure}

    In Fig.~\ref{fig:fig4}a), we plot the conditioned MFPT against the reference magnetization $m_0=m_r$, overlaying simulations and the theoretical approximation. We observe that for $r \neq 0$, a maximum in the MFPT appears at an $m_0 \neq -1$, which is counter-intuitive as one would naturally expect trajectories starting farthest from the target to take the longest time to arrive. 

    Our intuition fails here because we are observing a heavily filtered ensemble of trajectories. By conditioning the survival probability exclusively on reaching $m = 1$, we introduce a strong selection bias. The rare trajectories originating near $m = -1$ that successfully reach $m = 1$ are essentially ballistic, that is, they traverse the magnetization space rapidly, avoiding both the nearby absorbing boundary and the interruptions due to exogenous shocks. Conversely, for higher $r$, the maximum conditioned MFPT emerges in the intermediate region. Trajectories starting here are sufficiently far from $m = -1$ to avoid immediate absorption, allowing them to wander. This prolonged excursions subject them to multiple shocks that shift the system back to $m_0$. They survive long enough to be continuously returned, creating a cyclic trap that drastically inflates the mean time required to finally reach consensus at $m = 1$. The maximum appears for very small values of $r$ as we can see in Fig.~\ref{fig:fig4}b), and quickly reaches $m^*_0 = 0$ where it remains for increasing $r$. We observe that the simulations align satisfactorily for higher $r$ but, when the region of low rates is reached, our simulations disagree as we have very few samples that survive starting at $m_0$ very close to consensus.

    \section*{Exogenous events in complex networks}
    \label{sec:cn}
    
    Complex networks provide a more realistic representation of the interaction patterns found in real social systems. Significant attention has been devoted to unravel such patterns and how they impact the behavior of opinion dynamics models~\cite{newman2018networks, porter2016dynamical, carro2016noisy, suchecki2005voter, sood2008voter, sood2005voter, gleeson2013binary, pugliese2009heterogeneous}. Here, we follow this strand of research and show how to incorporate the exogenous events when the dynamics is defined on a network of agents. We stick to the voter model as a baseline dynamics.

    To obtain an analytical description, we follow the pair approximation proposed in~\cite{vazquez2008analytical}. Let us take a complex network of $N$ nodes (agents), with $\mu$ and $\mu_2$ as the first and second moments of the degree distribution. The degree-weighted magnetization $w$ is defined 
\begin{equation}
    w = \frac{1}{\mu N} \sum_i k_i s_i,
\end{equation}
    where $k_i$ is the degree of node $i$. The time-dependent degree-weighted magnetization probability density function $P(w,t)$ follows the Fokker-Planck equation
\begin{equation}
   \frac{\partial P(w,t') }{\partial t' } = \frac{\partial^2  }{\partial w^2 } \left( (1-w^2)P(w,t') \right),
   \label{eq:cn_fp}
\end{equation}
    where the normalized time $t'$ is
\begin{equation}
     t' = \frac{2\xi \mu_2}{\mu^2 N}t,
\end{equation} 
    and $\xi \equiv \frac{\mu-2}{2(\mu-1)}$ is a constant that depends only on the network topology.

    To exploit the discrete-time Markov chain framework to incorporate the exogenous shocks, we take the discretized version of Eq.~\ref{eq:cn_fp}. Thus, we recover the equivalent $\mathbf{P}$ matrix of the baseline model, now under the assumption that the agents interact in a network. Once available, the time-dependent probability mass function and the mean first-passage time are obtained in a similar manner to the previous section. Notice that, for the continuous case presented in the Supplementary Material, we obtained the solution of this differential equation in Eq.~\ref{eq:vm_sol}.

    We discretize the differential operator in Eq.~\ref{eq:cn_fp}, using the number of nodes $n$ with opinion $1$ for convenience. Thus,
\begin{multline}
   P(n,t+1) - P(n,t) = \frac{2\xi \mu_2 }{\mu^2N } \Big( (n-1)(N-n+1)P(n-1,t) \\
   +(n+1)(N-n-1)P(n+1,t) - 2n(N-n)P(n,t) \Big).
\end{multline}
    Note that since we use $m = 2n/N -1$ in Eq.~\ref{eq:cn_fp} the magnetization dependence disappears. This entails an important assumption, namely, the degree-weighted magnetization and the spin magnetization are approximated to be equal. Put otherwise, for each group of agents of degree $k$, the spin magnetization is equal and thus the same to the total magnetization. In doing this, we are assuming that there are no strong correlations between the degree of an agent and its opinion. We see that this approximation performs well at the level of description we are interested. 

    By changing the notation of the probability density function (PDF) $(P(n,t))$ to that of the probability mass function (PMF) $(a^{(t)}_n)$, we arrive at
\begin{equation}
  a_n^{(t+1)}  = \tilde p_{n-1}a_{n-1}^{(t)}  + \tilde p_{n+1}a_{n+1}^{(t)} + \left(1- 2\tilde p_{n}\right)a_n^{(t)},
\end{equation}
    where, defining $\theta \equiv n/N$ for convenience, the new rates are
\begin{equation}
    \label{eq:p_resc}
    \tilde p_{n} = \frac{ \mu_2(\mu-2)}{\mu^2(\mu-1) } \theta (1-\theta).
\end{equation}
    In the large-$N$ limit, this corresponds to a simple rescaling of the original rates by a factor that depends on the first and second moments of the degree distribution. 
    Thus, to obtain a description of the composite process on complex networks, we exploit the framework presented above, where now the new baseline matrix $\mathbf{P}$ has the entries of Eq.~\eqref{eq:p_resc}. The matrix $\mathbf{R}$ contains the information of the exogenous shocks and different choices are possible according to modeling needs.

\begin{figure}[t!]
    \centering
    \includegraphics[width=1\linewidth]{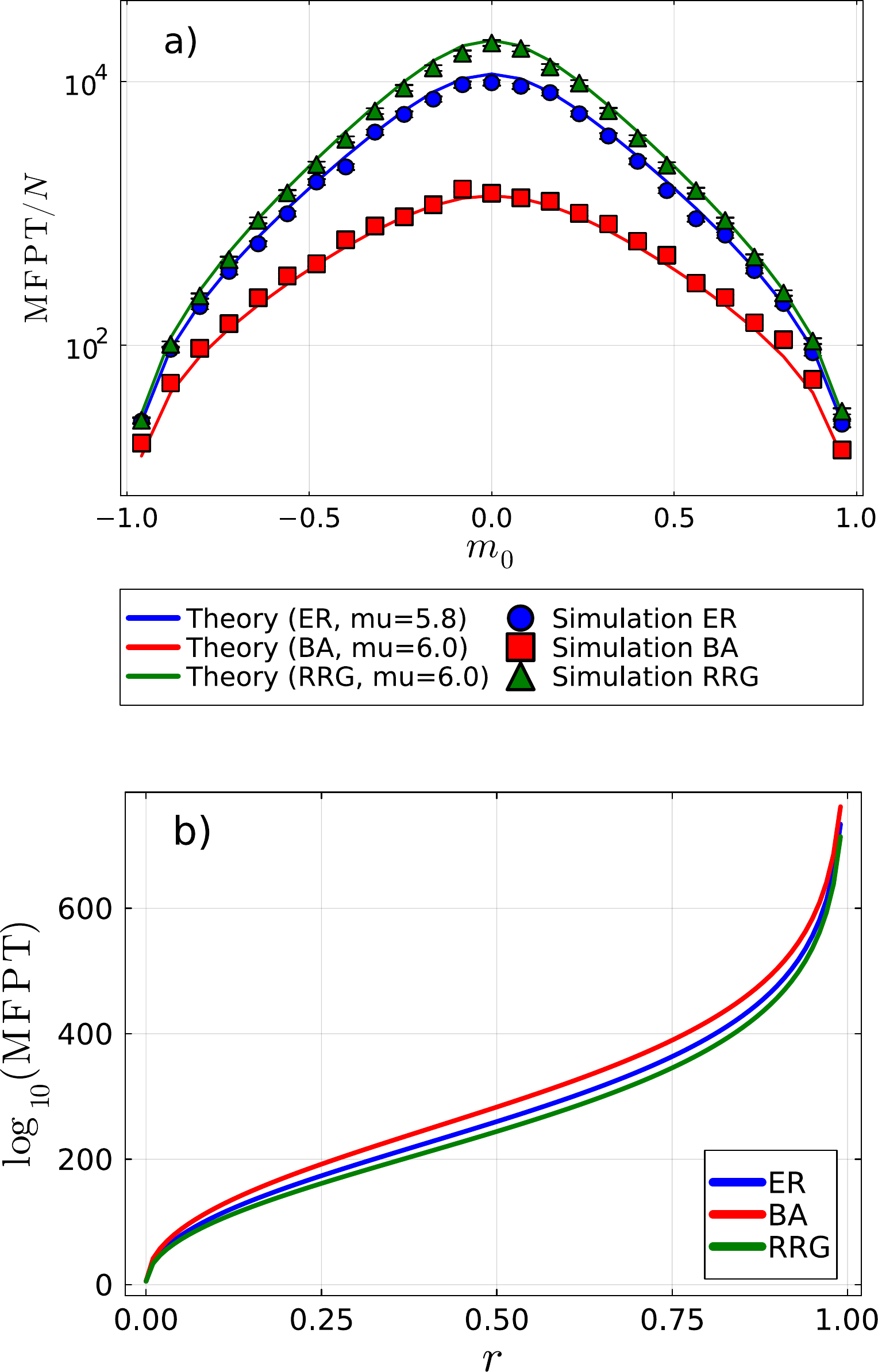}
    \caption{\textbf{Analysis of the voter model under the effect of exogenous shocks on networked topologies}. In a), the MFPT to consensus across different network topologies. Theoretical predictions from the pair approximation (lines) and Monte Carlo simulations (markers) for a random regular graphs (RRG), Erdős-Rényi (ER) networks, and Barabási-Albert (BA) networks. We consider networks with mean degree $\mu \approx 6$, $N = 500$ nodes, and a shock probability of $r = 0.0001$. In b), there is the behavior of the mean-first passage time at $m_0=0$ as a function of the shock rate $r$ for the same network ensembles.}
	\label{fig:fig5}
\end{figure}

    We validate the theory for the case of exogenous events that drive the system to a reference value of the magnetization $m_r$, i.e., the $\delta$-shocks employed above. In Fig.~\ref{fig:fig5}a), we show the comparison between simulations and the theoretical value of the mean first-passage time as a function of the reference magnetization $m_r=m_0$ and a moderate value of $r$. We have chosen three network ensembles with same mean degree $\mu$ but varying levels of heterogeneity in the connectivity ($\mu_2$): random regular graphs, Erdős-Rényi networks and a Barabási-Albert scale-free networks. In all cases, we see that the agreement is excellent. As in the all-to-all case, we verify in Fig.~\ref{fig:fig5}b) that reaching consensus is extremely unlikely when exogenous shocks are present: the mean-first passage time grows very fast with $r$. 

    We conclude this section by highlighting that studying the dynamics of exogenous shocks in complex networks opens the door to designing targeted interventions and implementation protocols that exploit structural information. Indeed, in a complete graph, an exogenous event that modifies the magnetization is straightforward because all nodes are structurally equivalent; that is, it does not matter which node changes its opinion. In contrast, in complex networks, the heterogeneity of agents' positions within the network (e.g., degree, centrality, etc.) implies that the same target magnetization $m_r$ after a shock can be achieved in many different ways. A natural way to incorporate this structural information is through the interface density, $\rho$, defined as the fraction of links connecting agents with different opinions. In general, $\rho$ is not uniquely determined by the magnetization but also depends on the specific assignment of opinions to the nodes. As a result, different shocks may produce different values of $\rho$ even when they yield the same target magnetization. Understanding the coupled dynamics of $m$ and $\rho$ lies beyond the scope of the present work and is left for future investigation.

    \section*{Exogenous shocks in the noisy voter model}
    \label{sec:noisy}

    As mentioned above, the mathematical framework to incorporate exogenous shocks developed here is general and can be readily extended to other opinion dynamics models. We support this claim by considering next the all-to-all version of the noisy voter model (NVM). 
    
    The NVM is defined such that there are two possible types of events: with probability $\eta$, an agent spontaneously changes opinion regardless of the opinion of its neighbors, while with the complementary probability $1-\eta$ a standard voter model occurs. Unlike the standard VM, the NVM is ergodic, as the consensus states are no longer absorbing given the spontaneous flips. As a function of $\eta$, there exist a modality transition in the stationary distribution: from a unimodal state at $m=0$ in the noise-dominated phase ($\eta > \eta_c$) to a bimodal state with maxmima at $m = \pm 1$ in the herding-dominated phase ($\eta < \eta_c$). At the transition point $\eta_c$, the stationary distribution is flat, meaning that all states are equiprobable. Our aim is then to unravel how this rich phenomenology is affected by the presence of exogenous shocks.
   
    To tackle this endeavor, we use the Markov chain theory with rates 
\begin{equation}
    \begin{aligned}
        \label{eq:rates_nvm}
        p^\uparrow_j &= (1-\theta) \left[ \eta + (1-\eta)\;\theta \right] \\
        p^\downarrow_j &=  \theta \left[ \eta + (1-\eta)(1-\theta) \right],
    \end{aligned}
\end{equation}
    where $p^\uparrow_j \equiv P(\Delta n(t) = 1 \,|\,n(t) = j)$ and $p^\downarrow_j\equiv  P(\Delta n(t) = -1 \,|\,n(t) = j)$. Note that, at odds with the standard voter model, the probability of increasing or decreasing the magnetization is not equal because the noisy updates create a drift toward $m = 0$. Given the rates, the discrete-time master equation of the baseline model, i.e., the NVM, can be written as
 \begin{equation}
  a^{(t+1)}_j =   a^{(t)}_{j+1} p^\downarrow_{j+1} + a^{(t)}_{j-1} p^\uparrow_{j-1} + a_{j}^{(t)} \left[ 1 - \left( p^\uparrow_j + p^\downarrow_j \right) \right].
 \end{equation}
    At this point, the workflow remains the same. We construct the modified $\mathbf{P}_r$ matrix from the rates~\eqref{eq:rates_nvm} and the matrix $\mathbf{R}$ containing information of the exogenous shocks. Then, the probability mass function of the composite process is simply given by
\begin{equation}
    \label{eq:baseline_matrix3}
    \mathbf{a}_r^{(t)} = \mathbf{P}_r^t \cdot \mathbf{a}_r^{(0)}.
\end{equation}
    Thus, obtaining both the time-dependence of the probability mass function and the stationary state of the system is, again, simply a matter of matrix exponentiation.

\begin{figure*}[t!]
    \centering
    \includegraphics[width=1\linewidth]{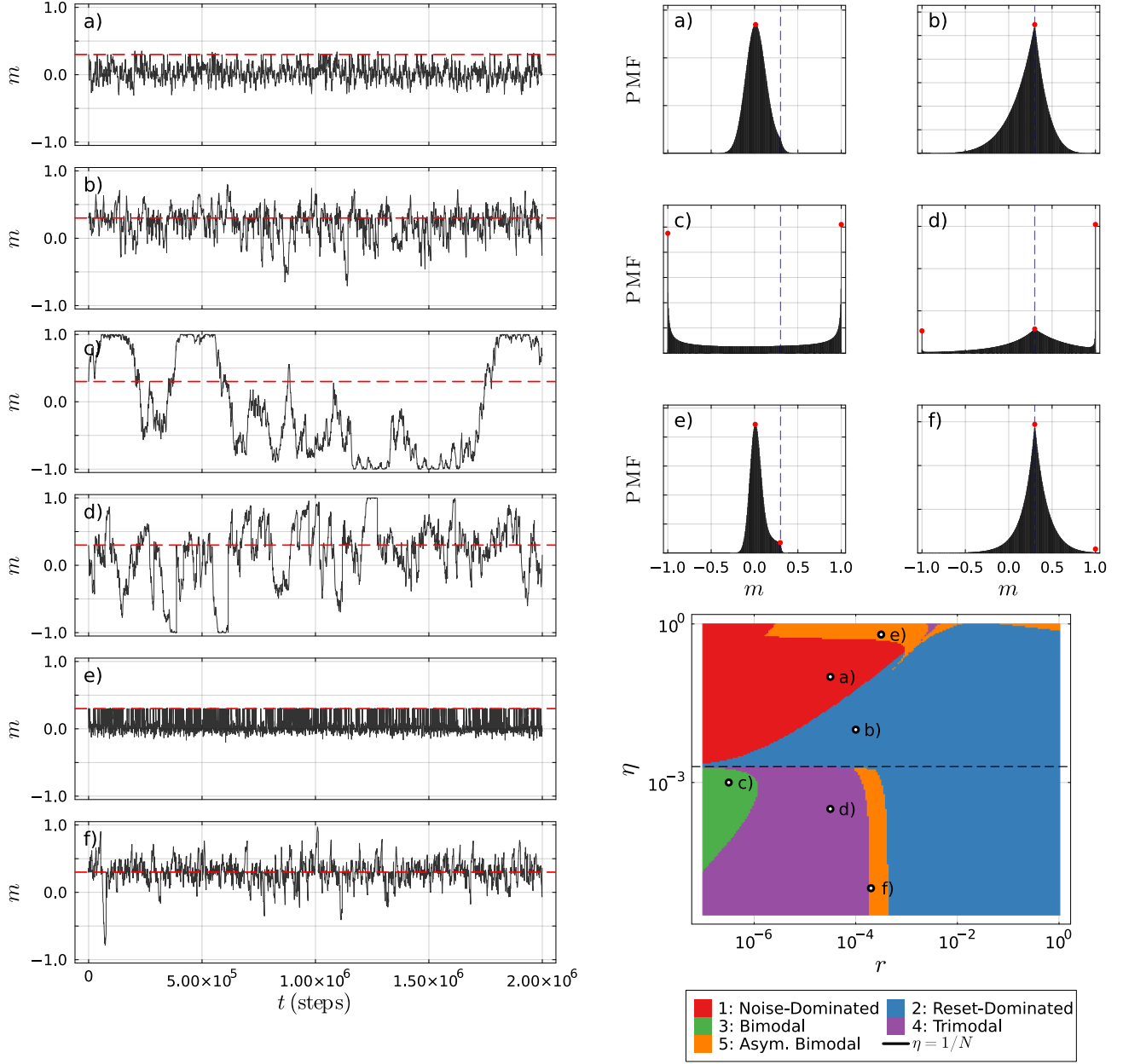}
    \caption{\textbf{Analysis of the noisy voter model under the exposure of exogenous shocks}. The left panels display Monte Carlo simulations of single trajectories of the magnetization of the noisy voter model with shocks. The horizontal dashed line mark the reference value $m_r$. Sharp vertical discontinuities in the trajectories indicate the presence of shocks and the subsequent jump to $m_r$. Each panel illustrates the regimes discussed in the main text: noise-dominated (a), shock-dominated (b), bimodal (c), trimodal (d), upper asymmetric bimodal (e) and lower asymmetric bimodal (f). On the top right part, we present the stationary probability mass function of the distinct dynamical regimes, obtained from averaging over the trajectory ensemble. The vertical dashed line corresponds to $m_r$, while red markers identify the local maxima. On the bottom right, we show the $(r, \eta)$-phase diagram. Here, the horizontal dashed line denotes the finite-size critical noise threshold $\eta_c = 1/N$ of the standard noisy voter model. The letters a), ..., f) correspond to the different phases mentioned above, and their position in the $(r, \eta)$-space specify, precisely, the values of the shock and noise rates used in the other panels. We consider a system size of $N = 500$, and $m_0 = m_r = 0.3$.}
	\label{fig:fig6}
\end{figure*}

    In the remaining, for illustrative purposes, we focus on the $\delta$-shocks again. A quick exploration of the dynamics for different values of noise $\eta$ and shock rates $r$ already hint to a much richer phenomenology than the baseline noisy voter model, with the emergence of new phases. In Fig.~\ref{fig:fig6} (left column), we show individual trajectories for particular combinations of $\eta$ and $r$, alongside their stationary probability mass functions (top right in Fig.~\ref{fig:fig6}). The selected $(\eta,r)$ pairs help us characterize the different phases, which we classify according to the number, location and global/local nature of the maxima of the PMF. This yields the following classification:
        
    \begin{itemize}
    \item[$\circ$] \textit{Noise-dominated phase:} There is a single peak at $m = 0$ due to the noise. That translates into the value of the magnetization fluctuating around $m = 0$, with exogenous shocks being rare and having very little impact.
    \item[$\circ$] \textit{Shock-dominated phase:} There is single peak at $m = m_r$ due to the exogenous events. In the individual trajectories, we observe fluctuations around $m_r$ with a noticeable bias toward $m = 0$ due to the low value of the shock rate $r$. That is, exogenous events dominate the dynamics but noise imbalances the fluctuations.
    \item[$\circ$] \textit{Bimodal phase:} There are two peaks at $m = \pm 1$ due to the normal VM dynamics bringing the system to consensus. In this phase, exogenous shocks are very rare and the system is allowed to explore the whole magnetization space spending some time in the consensus states before the noise or a shock takes the system out of consensus.
    \item[$\circ$] \textit{Trimodal phase:} Three peaks are present, at $m = \pm 1$ and $m = m_0$. The VM dynamics and the effects of exogenous shocks are visible together. We observe that the system has room to explore the whole magnetization space and even spend some time in consensus, but exogenous events have a more important role and condition the system to fluctuate more around $m_r$ while not in consensus.
    \item[$\circ$] \textit{Asymmetric bimodal phase:} This phase is characterized by two peaks, one at $m = \pm 1$ or $m = 0$ and the other at $m = m_r$. We further refine this phase classification by identifying the \textit{upper asymmetric bimodal phase}, where the noise exerts a strong drift toward $m = 0$ that is countered by very frequent exogenous events. This creates two clear maxima at $m = 0$ and $m_r$. We observe as well the \textit{lower asymmetric bimodal phase}, where there exist fluctuations around $m_r$ but some excursions end up in the $m = 1$ consensus state, creating a local maximum there.
\end{itemize}
    
    In Fig.~\ref{fig:fig6} we present the phase diagram as a function of $r$ and $\eta$, for the particular value $m_r = 0.3$. As can be appreciated, the boundaries between phases are distributed in a non-trivial way, with the same phase appearing in different regions of the $(\eta, r)$ values that are not adjacent. It is verified, as expected, that in the limit of $r \to 0$, the system behaves as the classic noisy voter model, displaying a modality transition exactly at $\eta = \eta_c = 1/N$~\cite{peralta2018stochastic}. Beyond this, the arrangement of the phases is highly dependent on the reference magnetization value $m_r$, highlighting the complex interplay between the properties of exogenous shocks and the baseline dynamics.

    \section*{Discussion}
    \label{sec:discussion}

    In this work, we have introduced a mathematical framework to incorporate large-scale exogenous shocks into the dynamics of opinion models that are defined through endogenous, local rules. Hence, we accommodate the possibility of modeling events that induce macroscopic changes in the system, disentangling them from the baseline dynamics. We have focused on the voter model as case study, being it a paradigmatic model for consensus formation. We have shown how to extract several quantities to characterize the non-equilibrium behavior of the model, such as the probability density function of the magnetization and first-passage statistics to consensus, under different modeling approaches. A discrete-time Markov chain formalism has provided a highly stable numerical framework to evaluate non-equilibrium quantities. Complementarily, a continuous-time description, more amenable to obtain closed-form expressions, has been laid out; see Appendix. Moreover, we have also tackled the case in which exogenous shocks influence agents interacting in a complex networks. By employing a pair approximation, we have analytically demonstrated that network heterogeneity scales the transition rates and that we can predict well the mean first-passage time across network topologies. Finally, to illustrate the generality of the approach, we have analyzed the all-to-all noisy voter model. In particular, we have identified new steady-state phases, such as the trimodal and the asymmetric bimodal regimes.
   
    While the analytical and numerical results establish a robust theoretical foundation, connecting these findings to empirical data represents a crucial next step. Indeed, in this first article we have explored simple, ad-hoc shock statistics. However, macroscopic opinion states could be quantified using longitudinal sentiment analysis or survey results. Then, the distribution of exogenous events $\psi(m_r)$ could be extracted from time series of relevant variables by identifying sudden, massive shifts in collective sentiment triggered, for instance, by breaking news, political scandals or coordinated misinformation campaigns. This is particularly relevant in the voter model, which in the past it was shown to be able to explain electoral data~\cite{fernandez2014voter}. Enriching it with large-scale exogenous shocks could provide explanatory power to even more complex scenarios. Another interesting avenue could be to perform experiments in order to verify whether the optimal shock rate $r^*$ identified here could exist in real populations, thus opening the door to control system in such a way that shocks could hinder or actively accelerate coordinated activities and collective decision-making. Recently, the voter model have made strides into explaining animal behavior, as well as robot swarms, so these experiments could be performed in a controlled manner in these settings. 

    From the modeling side, future work should aim to obtain better approximations for the composite process with shocks on complex networks, specially when the generalization to other baseline models is necessary. This includes, for instance, developing an approximated master equation~\cite{gleeson2013binary} or a stochastic pair approximation~\cite{peralta2020binary} versions of the treatment. Also, further investigation is required in the characterization of the effects of the distribution $\psi(m_r)$, for instance, by dropping the Poissonian assumption and extending to potentially correlated, power-law distributed exogenous events~\cite{karsai2018bursty, artime2017dynamics}. 

    More broadly, understanding collective behavior requires moving beyond the assumption that macroscopic dynamics emerge solely from the accumulation of local interactions: in many social, biological, and technological systems, rare and collective external perturbations can fundamentally reshape the trajectories of the system, and developing mathematical tools to describe their interplay with endogenous dynamics may therefore be essential for understanding, predicting, and ultimately steering the emergence of collective states.

    \section*{Appendix}
    \label{sec:SM}

    In this appendix, we complement the results introduced in the main text by developing a continuous-time framework to incorporate the large-scale exogenous shocks to baseline models. To be coherent with the content of the article, we particularize for the voter model case, but the description remains general to any one-dimensional stochastic process. 
    
    We highlight that the continuous-time description has some advantages and drawbacks with respect to the discrete-time one. As advantages, it seems more suitable to reach analytical, closed-form solutions from which one can extract simple asymptotic behaviors for the quantities of interest. Moreover, it is more flexible to accommodate arbitrary exogenous event distributions: while in the discrete-time one needs to work with matrices satisfying certain strict conditions, in the continuous-time case all the information is encoded into a single, parametrizable probability density function $\psi(m_r)$. As drawbacks, it helps knowing beforehand the solution of the Fokker-Planck equation of the baseline model, which is something that is not always available. More importantly, in our tests we find that the continuous-time case suffers of numerical instabilities that are difficult to circumvent in a general way. Our tests point to the fact these instabilities are not present at all in the discrete-time description. 
    \subsection{General solution for arbitrary exogenous shocks}
    \label{sec:continuous}

    We consider the voter model on an all-to-all network of $N$ nodes. As defined in Section~\ref{sec:vm}, we have changes of $\Delta t = 1/N$ and $\Delta m = 2/N$ at each step, but we take them as continuous, which is a reasonable approximation for large $N$. Then, the probability $G_0(m,t)$ of having a magnetization $m$ at time $t$ will be given by the following Fokker-Planck equation~\cite{vazquez2008analytical},
    \begin{equation}
       \frac{\partial G_0(m,t) }{\partial t } = \frac{1}{N} \frac{\partial^2  }{\partial m^2 } \left[(1-m^2)G_0(m,t)\right].
    \end{equation}
    This equation has known solution in terms of an infinite sum of Gegenbauer polynomials~\cite{mckane2007singular}
    \begin{widetext}
      \begin{equation}
        \begin{aligned}
        G_0(m,t|m_0) &= {}  \frac{1}{2}\delta(m-1)\left( 1+m_0\right)\left[1-\sum_{n=0}^{\infty}\frac{(1-m_0)(2n+3)}{(n+1)(n+2)}\,C_n^{(3/2)}(m_0)\,e^{-2(n+1)(n+2)t/N}\right] \\
        & + \frac{1}{2} \delta(m+1)\left( 1-m_0 \right)\left[1-\sum_{n=0}^{\infty}\frac{(1+m_0)(2n+3)}{(n+1)(n+2)}(-1)^n C_n^{(3/2)}(m_0)\,e^{-2(n+1)(n+2)t/N}\right] \\
        & + \frac{1}{2}\theta(m-1)\theta(m+1)(1-m_0^2)\sum_{n=0}^{\infty}\frac{(2n+3)}{(n+1)(n+2)}C_n^{(3/2)}(m_0)C_n^{(3/2)}(m)\,e^{-2(n+1)(n+2)t/N},
        \end{aligned} 
        \label{eq:vm_sol}
      \end{equation}
    \end{widetext}
    where $C_n^{(3/2)}(m_0)$ are the Gegenbauer polynomials of order $3/2$ and $m_0$ is the initial magnetization. Notice the delta terms at the consensus states $m = \pm 1$, reflecting the presence of absorbing states whose probability grows in time. In the steady state $t \to \infty$ only the delta terms remain and the probability in the region $m\in(-1,1)$ vanishes, indicating that consensus states are indeed absorbing. Large-scale exogenous events will break the absorbing nature of the consensus states, but the delta terms will be present as probability is accumulated there waiting for such an event.
    
    To mathematically introduce large-scale exogenous events, we only need to add two new terms to the Fokker-Planck equation~\cite{evans2020stochastic},
    \begin{equation}
        \begin{aligned}
            \label{eq:fpeq_exo}
            \frac{\partial P(m,t) }{\partial t } = & \frac{1}{N} \frac{\partial^2  }{\partial m^2 } \left[(1-m^2)P(m,t)\right] \\
            & - rP(m,t) + r\psi(m).
        \end{aligned}
    \end{equation}
    Here, the terms $-r P(m,t)$ and $r\psi(m)$ represent probability sinks and sources, respectively. At time $t$, an exogenous shock occurs at a rate $r$, so the system is taking from magnetization $m$ to a magnetization $m_r$ drawn from the distribution $\psi(m)$.   
    
    Eq.~\eqref{eq:fpeq_exo} has an exact solution as a function of the original solution $G_0(m,t)$ given by the last-renewal equation~\cite{evans2020stochastic},
    \begin{equation}
        \begin{aligned}
            P(m,t| & m_0) = e^{-rt} G_0(m,t|m_0) \\
            & + r \int_0^t \mathrm{d} s\; \int_{-1}^1 \mathrm{d} m_r\; e^{-rs}G_0(m,s|m_r)\psi(m_r).
       \label{eq:ren}
       \end{aligned}
    \end{equation}
    To solve Eq.~\eqref{eq:ren}, it suffices to perform an integral once $G_0$ is known, instead of solving a completely new differential equations. This is the approach we follow next, particularizing for different distributions $\psi(m_r)$. When it comes to interpret the steady state of the process with exogenous shocks, we note that it is a non-equilibrium steady state, since there is a constant flow of probability toward $m_0$ generated by the exogenous events, but this flow is perfectly countered by the outflow generated by the voter model diffusion.\\
    
    \subsection{Exogenous event probability distributions}

    In this section, we illustrate the validity of Eq.~\eqref{eq:ren} by particularizing for two exogenous event probability distribution $\psi(m_r)$. We focus on two extreme cases: a delta distribution $\psi(m_r) = \delta(m_r - m_0)$, in which the exogenous events take the magnetization to its initial value, and a uniform distribution $\psi(m_r) = 1/2$, in which the exogenous shocks take the magnetization to any value with equal probability. Note that the delta distribution is the continuous-time version of the case studied in the main text with the discrete-time approach. For the delta distribution, the exogenous events could take the system to any other value other than $m_0$, but this would only affect the transient. 
    
    We obtain the full time evolution, i.e., both the transient and steady-state regimes. For clarity, we only present the results and compare them with simulations in the steady state, where the last renewal equation simplifies to the integral term alone.
    
\begin{figure}[ht]
        \centering
        \includegraphics[width=1\linewidth]{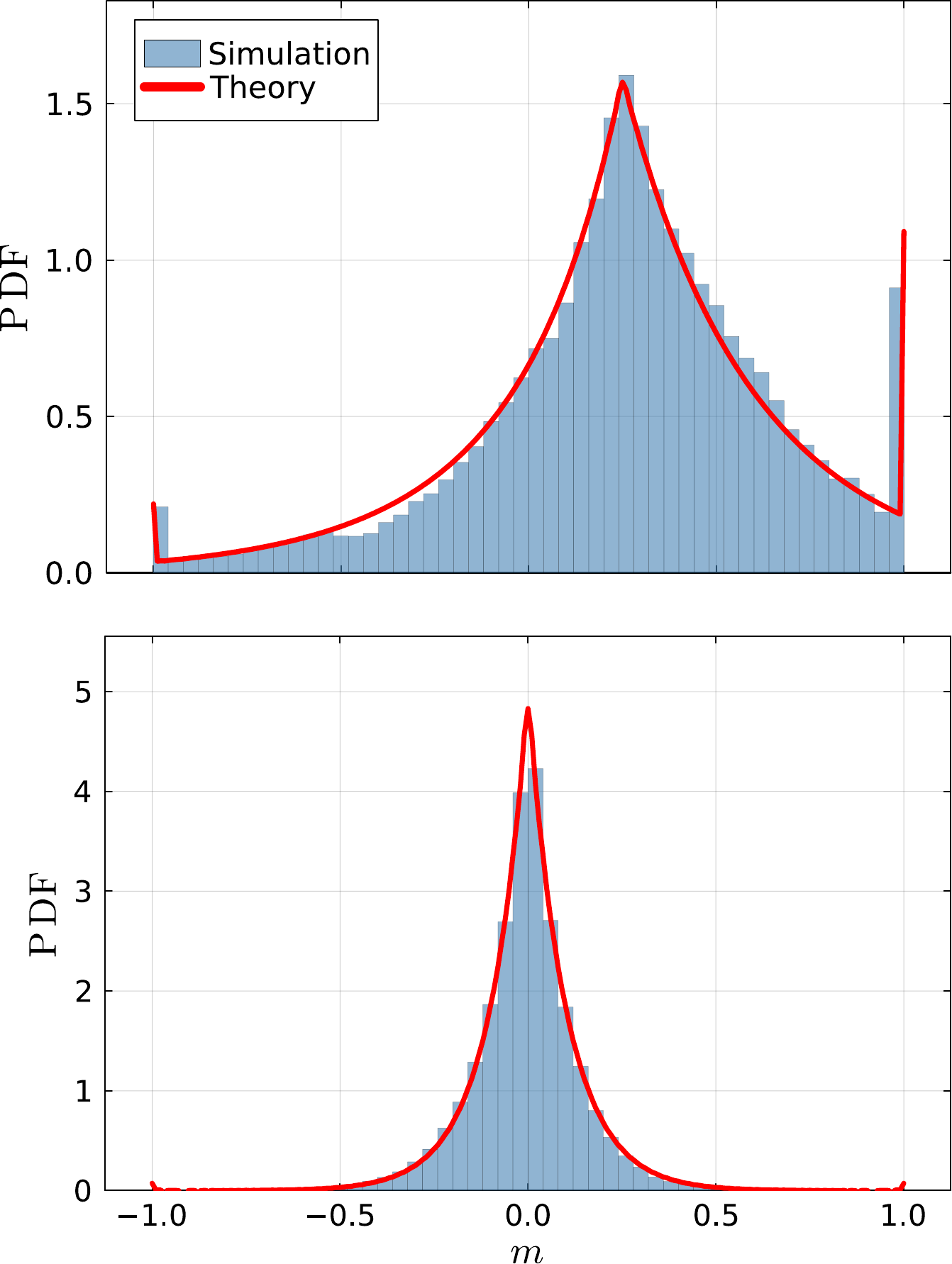}
        \caption{\textbf{Steady-state probability density function for delta shocks}. Analytical solution evaluated via Eq.~\ref{eq:del} (red line) compared with Monte Carlo simulations (bars). On the top, we consider $N = 1000$, rate $r = 0.01$ and $m_0 = 0.25$; on the bottom, $N = 1000$, rate $r = 1.0$ and $m_0 = 0$.}
        \label{fig:delta}
    \end{figure}

    For the delta exogenous events, we perform the integral in Eq.~\eqref{eq:fpeq_exo} with $\psi(m_r) = \delta(m_r-m_0)$. In the steady state, we obtain
    \begin{widetext}
      \begin{equation}
        \begin{aligned}
        P(m, t|m_0) &= \frac{1}{2} \delta(m-1)\left( 1+m_0\right)\left[1 - (1-m_0)\left(\sum_{n=0}^\infty \frac{(2n+3)}{(n+1)(n+2)}\frac{Nr}{(n+1)(n+2) +Nr}C_n^{(3/2)}(m_0)\right)\right] \\
        & + \frac{1}{2} \delta(m+1)\left( 1-m_0 \right)\left[1 - (1+m_0)\left(\sum_{n=0}^\infty \frac{(2n+3)(-1)^n}{(n+1)(n+2)} \frac{Nr}{(n+1)(n+2) +Nr}C_n^{(3/2)}(m_0)\right)\right] \\
        & +\frac{1}{2}\theta(m-1)\theta(m+1)(1-m_0^2) \sum_{n = 0}^\infty \frac{(2n+3)}{(n+1)(n+2)}\frac{Nr}{(n+1)(n+2) +Nr}C_n^{(3/2)}(m)C_n^{(3/2)}(m_0).
        \end{aligned}
        \label{eq:del}
      \end{equation}
    \end{widetext}
    In Fig.~\ref{fig:delta} we verify that the agreement between theory and simulations for different values of the parameter $r$.
    We see a characteristic peak at $m_0$ due to an accumulation of probability generated by the shocks. When the shock rate $r$ becomes higher, the consensus peaks diminish in height, indicating that high enough values $r$ inhibits consensus formation. 

    As per the case of exogenous shocks that equiprobably drive the system to any magnetization, i.e., $\psi(m_r) = 1/2$, we obtain
    \begin{equation}
        \begin{split}
          P(m|m_0) &= {}  \frac{1}{2} \left( 1 - \frac{Nr}{2+Nr} \right) \delta(m-1) \\
           & + \frac{1}{2} \left( 1 - \frac{Nr}{2+Nr} \right) \delta(m+1) \\
           & + \frac{1}{2}\frac{Nr}{2+Nr} \theta(1-m)\theta(m+1).
        \end{split}
        \label{eq:uni}
    \end{equation}
    This is a much simpler expression as we have integrated out the Gegenbauer polynomials. 
    
\begin{figure}[ht]
        \centering
        \includegraphics[width=1\linewidth]{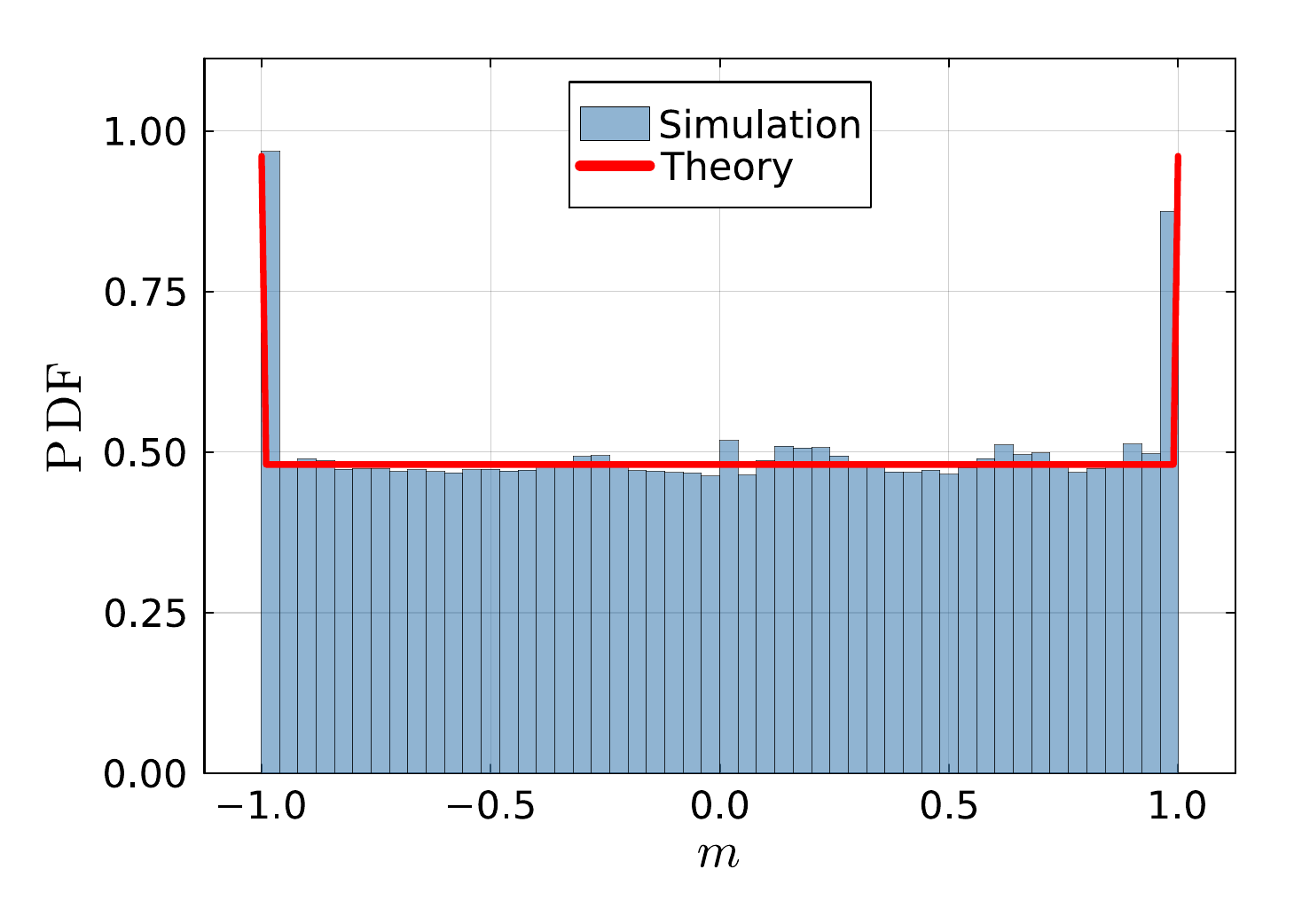}
        \caption{\textbf{Steady state PDF for uniform shock}. Analytical solution evaluated via Eq.~\ref{eq:uni} (red line) compared with Monte Carlo simulations (bars) for a system under a uniform event distribution $\psi(m_r) = 1/2$. We consider a system size of $N = 500$, with rate $r = 0.1$ and $m_0 = 0$.}
        \label{fig:uni}
    \end{figure}

    In Fig.~\ref{fig:uni}, we observe the effects of this type of exogenous shock.
    Contrary to the delta distribution, in this case, there is no preferred magnetization when an exogenous shock occurs. At sufficiently low $r$, the consensus states are significantly populated, resulting in two symmetric peaks at $m = \pm 1$. However, as $r$ increases, the peaks become progressively lower, since the voter dynamics, and hence the absorbing nature of the consensus states, turn out less relevant.
    
    \subsection{First-passage time description of the continuous-time process}
    
    The probability distribution $P(m,t|m_0)$ allows us to characterize additional quantities that are useful for assessing the effect of exogenous events, such as the first-passage time to consensus. Once again, we can use renewal theory to obtain the continuous FPT distribution provided we know it for the shock-free case. 
    
    In general, the FPT distribution for a stochastic process departing from $m_0$ satisfies~\cite{redner2001guide}
    \begin{equation}
       f(m_0, t) = - \frac{\partial S(m_0,t) }{\partial t },
    \end{equation}
    where $S(m_0,t)$ is the survival probability, i.e., for the voter model, the probability of not reaching consensus up to time $t$ having departed from $m_0$. When exogenous shocks are present, the survival probability of the process is given by the last renewal equation~\cite{evans2020stochastic}
    \begin{multline}
       S(m_0, t) = e^{-rt} S_0(m_0,t) \\
       + r \int_0^t e^{-r\tau} S_0(m_0, \tau)S(m_0, t-\tau)\; \mathrm{d} \tau,
       \label{eq:fpt_ren}
    \end{multline}
    where $S_0(m_0,t)$ is the survival probability of the shock-free case. Eq.~\eqref{eq:fpt_ren} cannot be easily solved just by doing an integral, since the unknown inside the integral. Yet, exploiting the convolution form of the equation, we employ the Laplace transform
    \begin{equation}
       \tilde S(m_0,s) \equiv \int_0^\infty \mathrm{d} t \; e^{-st}S(m_0,t).
    \end{equation}
    Thus, Laplace-transforming Eq.~\eqref{eq:fpt_ren},
    \begin{equation}
       \tilde S(m_0,s) = \tilde S_0(m_0,r+s) + r \tilde S_0(m_0,r+s)\tilde S(m_0,s).
    \end{equation}
    Isolating the transformed survival probability as a function of the transformed shock-free survival probability yields
    \begin{equation}
       \tilde S(m_0, s) = \frac{\tilde S_0(m_0,r+s)}{1-r \tilde S_0(m_0, r+s)} .
       \label{eq:FPT}
    \end{equation}

\begin{figure}[ht]
        \centering
        \includegraphics[width=\linewidth]{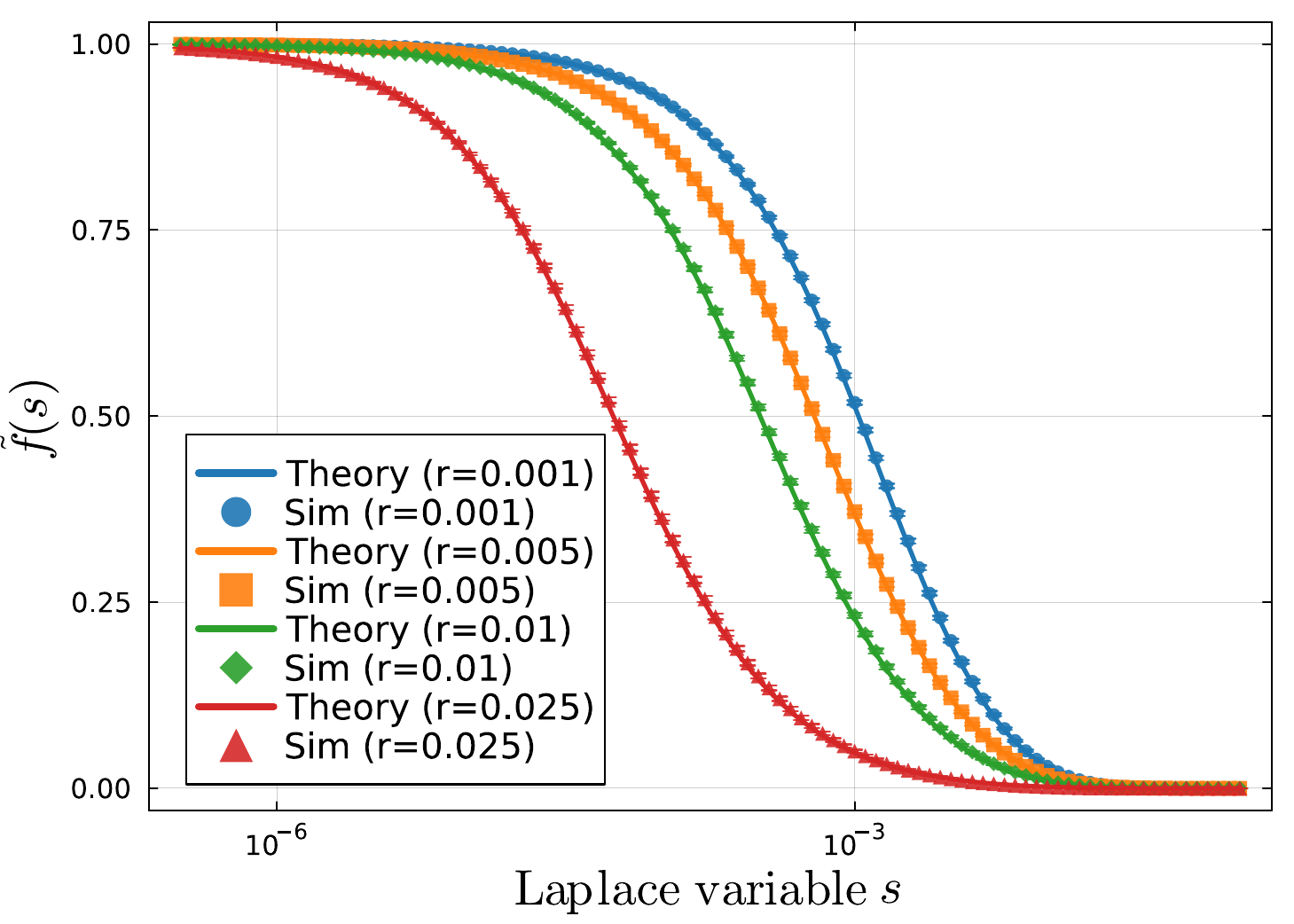}
        \caption{\textbf{Laplace transform of the first-passage distribution to consensus}. Analytical predictions computed via Eq.~(\ref{eq:FPT}) (solid lines) compared with Monte Carlo simulation data (symbols). We consider a system size of $N = 1000$, and $m_0 = 0$.}
        \label{fig:fpt}
    \end{figure}

    To apply the results derived above, we need $S_0(m_0,t)$, which is just 1 minus the probability to have reached consensus at time $t$. The latter is simply the sum of the coefficients associated to the terms $\delta(m-1)$ and $\delta(m+1)$ in Eq.~(\ref{eq:vm_sol}). Thus,
    \begin{widetext}
      \begin{equation}
        \begin{aligned}
       S_0(m_0, t) &= {} 1- \frac{1}{2}\left( 1+m_0\right)\left[1-\sum_{n=0}^{\infty}\frac{(1-m_0)(2n+3)}{(n+1)(n+2)}\,C_n^{(3/2)}(m_0)\,e^{-2(n+1)(n+2)t/N}\right] \\
        & - \frac{1}{2} \left( 1-m_0 \right)\left[1-\sum_{n=0}^{\infty}\frac{(1+m_0)(2n+3)}{(n+1)(n+2)}(-1)^n C_n^{(3/2)}(m_0)\,e^{-2(n+1)(n+2)t/N}\right].
        \end{aligned} 
      \end{equation}
    \end{widetext}
    Notice how in the multiplying term of each bracket the $1/2$ cancels the $1$ and the $m_0$ cancels each other, and in the sums we are left with only the even terms due to the alternating sign $(-1)^n$ in the second bracket, so we can simplify the expression and obtain,
    \begin{multline}
       S_0(m_0, t) = (1-m_0^2) \sum_{l = 0}^\infty  \frac{(4l+3)}{(2l+1)(2l+2)} \\
       \cdot C_{2l}^{(3/2)}(m_0) e^{-\frac{1}{N}(2l+1)(2l+2)t}.
    \end{multline}
    The Laplace transform is, then,
    \begin{multline}
       \tilde S_0(m_0, s) = (1-m_0^2) \sum_{l = 0}^\infty \frac{(4l+3)}{(2l+1)(2l+2)}\\
       \cdot C_{2l}^{(3/2)}(m_0) \frac{N}{Ns+(2l+1)(2l+2)}
    \end{multline}
    Introducing this into Eq.~(\ref{eq:FPT}) gives the FPT distribution in the Laplace space. This expression is not readily invertible, so to compare with simulations we take the simulation data to Laplace space, as can be seen in Fig.~\ref{fig:fpt}.
    Finally, we can see a good agreement between the theoretical results and the stochastic simulations across different shock rates $r$. Note that for $s\to 0$ we have that $\tilde f(s) \to 1$, which accounts for the FPT distribution being normalized. At this point, one could numerically anti-transform $\tilde f(s)$ to readily obtain the FPT distribution.

    \section*{Data Availability Statements}
    The data that supports the findings of this study are available within the article.
    
    \section*{Acknowledgements}

    GAH acknowledge the UBICS for granting a Beca de Col·laboració, during which this work was developed. OA acknowledge support from Ministerio de Ciencia, Innovación y Universidades, project PID2024-158120NB-C22.
    
    \bibliographystyle{unsrt}
    \bibliography{bibliography}


\end{document}